\documentclass{iopart}

\usepackage{graphicx}           % for eps figures

\begin{document}

\title[Entanglement between static and flying qubits in a nanotube]{Entanglement between static and flying qubits in a semiconducting carbon nanotube}

\author{D~Gunlycke$^1$, J~H~Jefferson$^2$, T~Rejec$^3$, A~Ram\v{s}ak$^4$, D~G~Pettifor$^1$ and G~A~D~Briggs$^1$}

\address{$^1$ Department of Materials, University of Oxford, Parks Road, Oxford, OX1~3PH, UK}

\address{$^2$ QinetiQ, St. Andrews Road, Malvern, WR14~3PS, UK}

\address{$^3$ Department of Physics, Ben-Gurion University of the Negev, Beer Sheva 84105, Israel}

\address{$^4$ Faculty of Mathematics and Physics, University of Ljubljana, J. Stefan Institute, 1000 Ljubljana, Slovenia}

\begin{abstract}
Entanglement can be generated by two electrons in a spin-zero state on a semiconducting single-walled carbon nanotube.  The two electrons, one weakly bound in a shallow well in the conduction band, and the other injected into the conduction band, are coupled by the Coulomb interaction.  Both transmission and entanglement are dependent on the well characteristics, which can be controlled by a local gate, and on the kinetic energy of the injected electron.  Regimes with different degrees of electron correlation exhibit full or partial entanglement.  In the latter case, the maximum entanglement can be estimated as a function of width and separation of a pair of singlet-triplet resonances.
\end{abstract}


\submitto{\JPCM}

%\maketitle
%\tableofcontents

\section{Introduction}
\label{sec:1}

Entanglement is a unique resource of quantum information processing (QIP), making possible the exploitation of the full range of quantum states and their associate manipulation.  Endohedral fullerenes in single-walled carbon nanotubes are candidates for the physical embodiment of QIP \cite{Arda03,Arda05}.  Carbon nanotubes have long decoherence lengths and exhibit ballistic transport near the quantum limit \cite{Whit98,Lian01,Kong01}.  Their cylindrical geometry permits filling of various species, and their special energy dispersions include electron-hole symmetry \cite{Jari04}.  Local gating of carbon nanotubes can be used to create quantum dots \cite{Maso04} and spin filters \cite{Gunl05}.  It has also been shown that two ballistic electrons scattering against a magnetic impurity can generate entanglement \cite{Cost01}.

Entanglement can be generated between two qubits, one static and one flying.  Flying qubits in QC serve the purpose of an information bus.  DiVincenzo extended his original five requirements \cite{DiVi96} of QC by including two further desirable conditions about flying qubits \cite{DiVi00}.  The qubits in our model are embodied by electron spin with respect to a certain axis.  The problem is analogous to treating the collision of an electron with a hydrogen atom, with due regard to Fermi statistics \cite{Oppe28,Mott30}.  In our model one of the electrons is injected into the conduction band of a semiconducting nanotube from an electrode, and the other electron is confined in a quantum well in the same band.  The well is located at a section of the nanotube somewhere in-between the electrodes and may be created by local gates or by some other effect causing electrostatic modulation.  Due to the high confinement in the transverse directions, the well has discrete energy levels and in this sense is a quantum dot.  The model is inspired by a similar study where a two-dimensional electron gas (2DEG) was used \cite{Jeff05}.  Incidentally, research on 2DEGs also suggests that unintentional shallow wells of the kind described could be responsible for the 0.7 anomaly in conductance experiments \cite{Reje00,Reje03}.  The dynamics in our model are governed by a long-range repulsive Coulomb interaction and a relatively short-range attractive well potential.  The Coulomb interaction was also exploited to generate entanglement between two electrons in another 2DEG study \cite{Sara04}.  Adding the potential contributions from the Coulomb interaction and the well forms a resonant structure with either two or three barriers depending on the choice of parameters.  We solve the scattering problem using $\pi$-orbital envelope function theory.  The reflection and transmission coefficients in the complete two-electron problem depend on the spin configuration of the two electrons.  This is due to the spin exchange that occurs while the two electrons have a wavefunction overlap.  The spin exchange generates entanglement shared between the two electrons.  We study the dependence of this entanglement on kinetic energy of the injected electron for different wells and show that there are three distinctly different pictures depending on electron correlation.

There are six sections in this paper.  The next section introduces some further detail of our model.  It formulates the basic requirements and presents a study of the well parameters.  \Sref{sec:3} covers single-electron approaches based on different Hartree approximations and in the subsequent section, we solve the complete two-electron problem.  \Sref{sec:5} discusses the entanglement generated in the scattering process, and finally, the results are then concluded in \sref{sec:6}.

\section{The model}
\label{sec:2}

\subsection{Preparations and requirements}

\begin{figure}
  \begin{center}
    \includegraphics[scale=0.5]{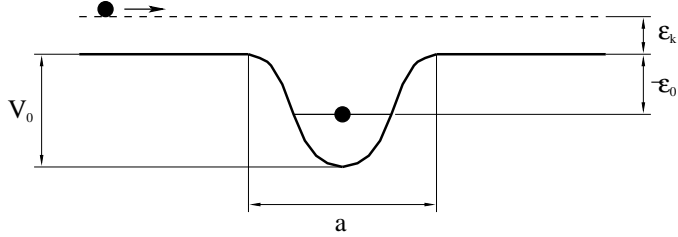}
  \end{center}
  \caption{One electron with energy $\varepsilon_0$ is located in the ground state of a shallow well.  The well has depth $V_0$ and total width $a$.  A second electron is injected into the conduction band with energy $\varepsilon_k$.}
  \label{fig:1}
\end{figure}

Before the entanglement process begins, the system is prepared with one of the electrons bound in a shallow well in the conduction band of the semiconducting nanotube.  The first step in obtaining this state is to apply a local external potential using either a narrow gate segment of the type used to fabricate carbon-nanotube field-effect transistors \cite{Wind03} or a narrow split gate \cite{Robi03}.  A single electron is then introduced into the nanotube using a turnstile injector and the system is allowed to relax to its ground state with the electron occupying the quantum well.  For the purpose of these calculations, the potential induced on the nanotube by the external gate is chosen to be the one-dimensional potential energy
\begin{equation}
  \label{eq:1}
  V(x) = \left\{
  \begin{array}{ll}
    V_0~\cos^2\left(\frac{\pi x}{a}\right)\quad & |x|\leq a/2,
    \\
    0 & |x|>a/2,
  \end{array}
  \right.
\end{equation}
where $x$ is the spatial coordinate along the direction of the nanotube axis, $V_0$ is the potential energy at the potential minimum, and $a$ is the total width of the well (\fref{fig:1}).
To avoid confusion with other potential energies to be discussed later, we refer to this well as the applied potential.  For a certain width $a$, the parameter $V_0$ is chosen to give a well shallow enough to accommodate one and only one electron and deep enough so that a second incident electron will not cause ionisation.  The bound electron occupies the ground state level which has energy $\varepsilon_0$.  Although the bound state can, in principle, accommodate two electrons of opposite spin, the Coulomb energy $U$, associated with adding a second electron in the dot, pushes the two-electron state into the continuum, provided that the well is sufficiently shallow.  This gives rise to a resonant state for the second electron with
\begin{equation}
  \label{eq:2}
  \varepsilon_0 + U \ge 0.
\end{equation}
The second electron is injected into the nanotube through a turnstile in conjuction with a spin filter.  The kinetic energy of the injected electron, $\varepsilon_k$, may be changed via gates but must be restricted to $\varepsilon_k < |\varepsilon_0|$ to avoid ionisation of the bound electron.  If the energy of the incident electron is tuned to the lowest resonance energy, we obtain the approximate relation $\varepsilon_k \sim \varepsilon_0 + U$.  Combining the two relations gives the following ionisation condition
\begin{equation}
  \label{eq:3}
  2\varepsilon_0 + U \le 0.
\end{equation}
We restrict the kinetic energy and well parameters to fulfil equations \eref{eq:2} and \eref{eq:3} to ensure single occupation of the well for asymptotic states.

\subsection{Parameter study}

\begin{figure}
  \begin{center}
    \includegraphics[scale=0.5]{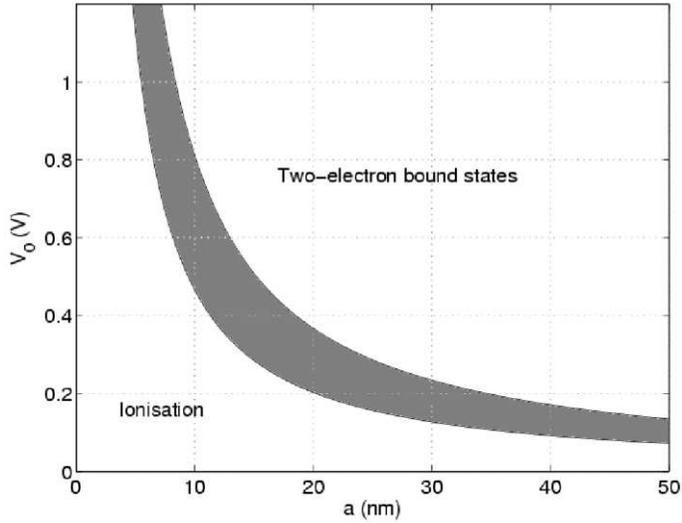}
  \end{center}
  \caption{Dependence of charge states in the quantum dot on the depth and the width of the well.  The parameters are chosen to lie in the shaded area, which gives single-electron occupation asymptotically.}
  \label{fig:2}
\end{figure}

The scattering of the propagating electron from the bound electron in a single-walled nanotube may be described accurately by an effective single-band model for the lowest conduction band, derived from envelope function theory combined with either $\pi$-orbital $\vec k\cdot\vec p$ or orthogonal tight-binding theory.  Within the framework of the effective mass approximation, both these theories yield the same result which, at the one-electron level, is an equation identical to the time-independent Schr\"odinger equation
\begin{equation}
  \label{eq:4}
  H\phi = E\phi.
\end{equation}
$H$ is the single-electron Hamiltonian
\begin{equation}
  \label{eq:5}
  H = -\frac{\hbar^2}{2m^*}\frac{d^2}{dx^2} + V(x),
\end{equation}
where $m^*=E_g/2v_F^2$ is the effective mass, with $v_F$ the fermi velocity of a single graphene sheet, and $V(x)$ is the effective one-electron potential energy in equation \eref{eq:1}.  The bound states of the quantum dot, with boundary conditions
\begin{equation}
  \label{eq:6}
  \lim_{x\rightarrow \pm \infty}\phi(x) = 0,
\end{equation}
were determined using the finite difference method.  We computed the range of the parameters $V_0$ and $a$ which gives a single bound state with energy $\varepsilon_0$ and corresponding wavefunction $\phi_0(x)$.  The eigensolutions may be obtained approximately by analogy with a parabolic potential,
\begin{equation}
  \label{eq:7}
  V(x) = V_0\left[1-\left(\frac{\pi x}{a}\right)^2\right]+\Or\left(x^4\right).
\end{equation}
For the parabolic potential the bound state energies are
\begin{equation}
  \label{eq:8}
  E_n=V_0 + \hbar\pi\sqrt{\frac{2|V_0|}{m^*a^2}}\left(n+\frac{1}{2}\right).
\end{equation}
With two electrons in the dot, we can estimate their Coulomb energy to be of order $U\sim e^2/4\pi\epsilon_0d$, where the effective electron separation is crudely approximated by $d\sim a/4$, using the axial dielectric constant $\epsilon_x=1$, in accordance with \cite{Leon02}.  Using these estimates and equations \eref{eq:7} and \eref{eq:8} allow us to rewrite conditions \eref{eq:2} and \eref{eq:3} in the form,
\begin{equation}
  \label{eq:9}
  0 \le s \le 1,
\end{equation}
where
\begin{equation}
  \label{eq:10}
  s \equiv \frac{2|\varepsilon_0|}{U} \approx \frac{2\pi\epsilon_0}{e^2}\left(V_0a-\frac{\hbar\pi v_F}{\sqrt{E_g}}\sqrt{V_0}\right)-1.
\end{equation}
As a specific example, we choose a $(19,0)$-nanotube which has band gap energy $E_g\approx 526$~meV and Fermi velocity $v_F\approx~7.25\cdot 10^5$~m~s$^{-1}$.  The parameter space for the quantum dot geometry is shown in \fref{fig:2}.
Parameters below the strip lead to a regime with finite probability of ionisation ($s \le 0$), whereas parameters above the strip can accommodate two  bound electrons ($s \ge 1$).  We choose parameters in the range between these two regimes, ensuring a single bound electron for the asymptotic states.

\section{Single-electron approximations}
\label{sec:3}

\subsection{The frozen Hartree approximation}

We impose the constraint that one electron remains bound in the dot in the one-electron state $\phi_0$ and consider the scattering of the second injected electron, taking into account the Coulomb repulsion between the two.  Although this frozen Hartree approximation is rather severe, it illustrates the origin of quasi-bound state resonances and also gives insight into where and why such an approximation breaks down.  The fixed charge distribution of the bound electron appears as a point charge at large distances with $1/r$-dependence whereas inside the well approaching the centre, there is competition between increasing Coulomb energy and decreasing potential energy due to the well.  To derive a quantitative expression for the effective potential we start with the Coulomb potential between the two electrons in the approximate form,
\begin{equation}
  \label{eq:11}
  V_C(x_1,x_2) = \frac{e^2}{4\pi\epsilon_0\sqrt{(x_1-x_2)^2+\lambda^2}},
\end{equation}
where the parameter $\lambda$ is the effective distance between the two electrons when $x_1=x_2$.  To estimate $\lambda$, we start from the unscreened three-dimensional Coulomb expression
\begin{equation}
  \label{eq:12}
  V_C(\vec r_1,\vec r_2) = \frac{e^2}{4\pi\epsilon_0\sqrt{(x_1-x_2)^2+R_1^2+R_2^2-2R_1R_2\cos\alpha}}.
\end{equation}
and integrate over the transverse modes around the circumference and radially.  For $x_1=x_2$ this gives directly,
\begin{equation}
  \label{eq:13}
  \frac{1}{\lambda} \equiv \frac{1}{2\pi}\int_0^{2\pi}\frac{\rmd\alpha}{2\sqrt{(R-\Delta R)^2\sin^2(\alpha/2)+(\Delta R)^2}}.
\end{equation}
where $R\approx 0.75$~nm~is the nanotube radius and $\Delta R\equiv 5a_0/Z_{\mathrm{eff}}\approx 0.0814$~nm~is the radial extent of a $\pi$-electrons.  For $x_1\ne x_2$, equation \eref{eq:11} remains an excellent approximation and is clearly correct asymptotically.

\begin{figure}
  \begin{center}
    \includegraphics[scale=0.5]{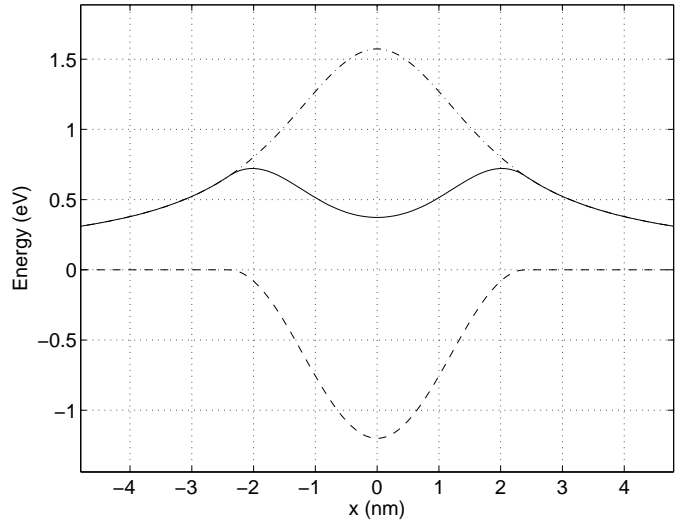}
  \end{center}
  \caption{Effective potential acting on the free electron in the frozen Hartree picture.  The dashed curve illustrates the applied potential, where $V_0=1.2$~eV and $a=4.8$~nm.  The dash-dotted curve represents the potential from the Coulomb repulsion to the fixed bound electron in the quantum dot and the solid line shows the total potential.}
  \label{fig:3}
\end{figure}

\begin{figure}
  \begin{center}
    \includegraphics[scale=0.5]{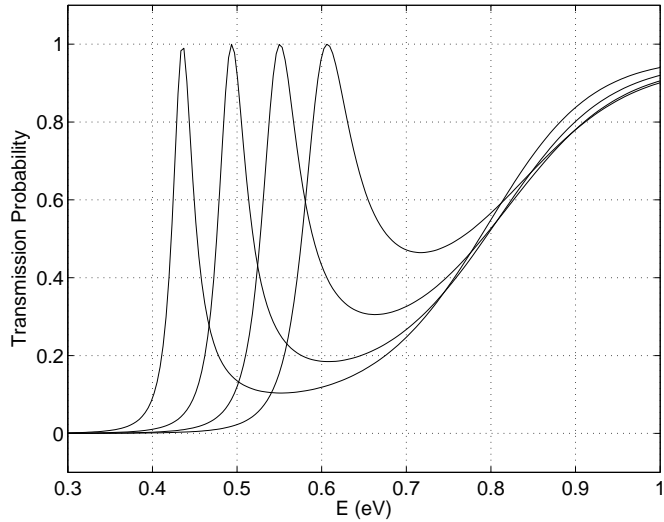}
  \end{center}
  \caption{Transmission probability of the free electron as a function of its kinetic energy, calculated within the frozen Hartree picture.  The curves with resonances from left to right are calculated using the depth of the well $V_0=(1.5,1.4,1.3,1.2)$~eV, in that order.  The width of the well is $a=4.8$~nm.}
  \label{fig:4}
\end{figure}

\begin{figure}
  \begin{center}
    \includegraphics[scale=0.5]{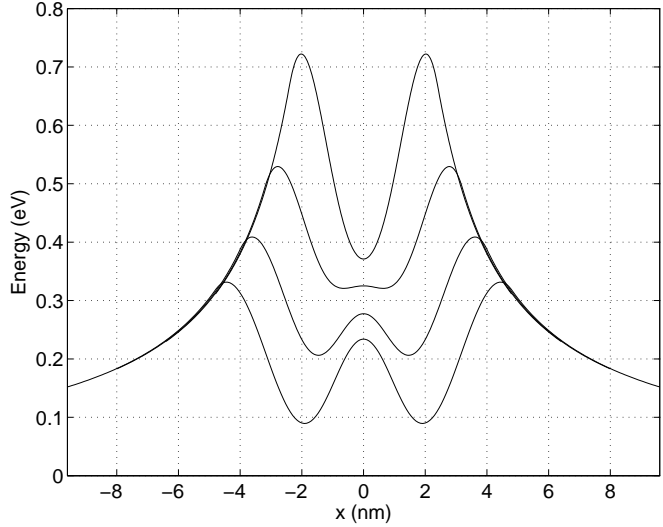}
  \end{center}
  \caption{Effective potentials operating on the free electron in the frozen Hartree picture.  The traces from top to bottom are calculated using the applied well widths $a=(4.8,6.4,8.0,9.6)$~nm, while the depth of the applied well has been kept constant at $V_0=1.2$~eV.  The frozen Hartree approximation requires a single well-defined well.  This is satisfied for the narrowest well but not for the wider wells where the approximation consequently fails.}
  \label{fig:5}
\end{figure}

The effective frozen Hartree Coulomb potential energy is the expectation value of equation \eref{eq:11} where the position of the bound electron is given by its envelope function $\phi_0(x)$;
\begin{equation}
  \label{eq:14}
  V_C(x) = \int_{-\infty}^{\infty}\frac{e^2}{4\pi\epsilon_0\sqrt{(x-x')^2+\lambda^2}}~|\phi_0(x')|^2~\rmd x'.
\end{equation}
Summation of the two potential contributions gives the total effective potential experienced by the propagating electron,
\begin{equation}
  \label{eq:15}
  V_{\mathrm{eff}}(x) = V(x) + V_C(x),
\end{equation}
which is shown in \fref{fig:3} together with the potentials in equations \eref{eq:1} and \eref{eq:14}.
The potential well in this example is narrow with $a=4.8$~nm, which gives a width for the ground-state wavefunction in the well of order the nanotube diameter.  Indeed this may be realized in practise by using a metallic nanotube itself to form the gate.  In this regime, the effective potential has the shape of a double barrier structure.  If the well parameters are chosen such that $0<s<1$, then the structure contains at least one resonance.  \Fref{fig:4} shows this resonance for various depths of the well ranging from $V_0 = 1.5$~eV to $V_0 = 1.2$~eV.
The curve with the rightmost resonance is calculated for the same parameters as in \fref{fig:3}, and in this case the resonant tunnelling occurs at $\epsilon_k\approx 605$~meV for the incident electron.  In \fref{fig:4}, we also see that the resonances move to lower energy and become narrower as the well depth increases, consistent with the decreasing tunnel probability through a single barrier.  When the well widths become larger, the frozen Hartree approximation starts to break down, as can be understood from \fref{fig:5} in which the effective potential gradually changes shape with increasing well width, eventually developing a double-well structure.
Since the two electrons can lower their mutual energy by occupying the two wells in the resonance state, we see that this is inconsistent with the frozen Hartree approximation.  Therefore, the approximation works best in the weak-correlation regime when the width of the well at the resonance energy is small or at least comparable to the Bohr radius $a_{NT}=(m/m^*)\epsilon_x a_0\approx 0.6$~nm.

\subsection{The Hartree approximation}

\begin{figure}
  \begin{center}
    \includegraphics[scale=0.5]{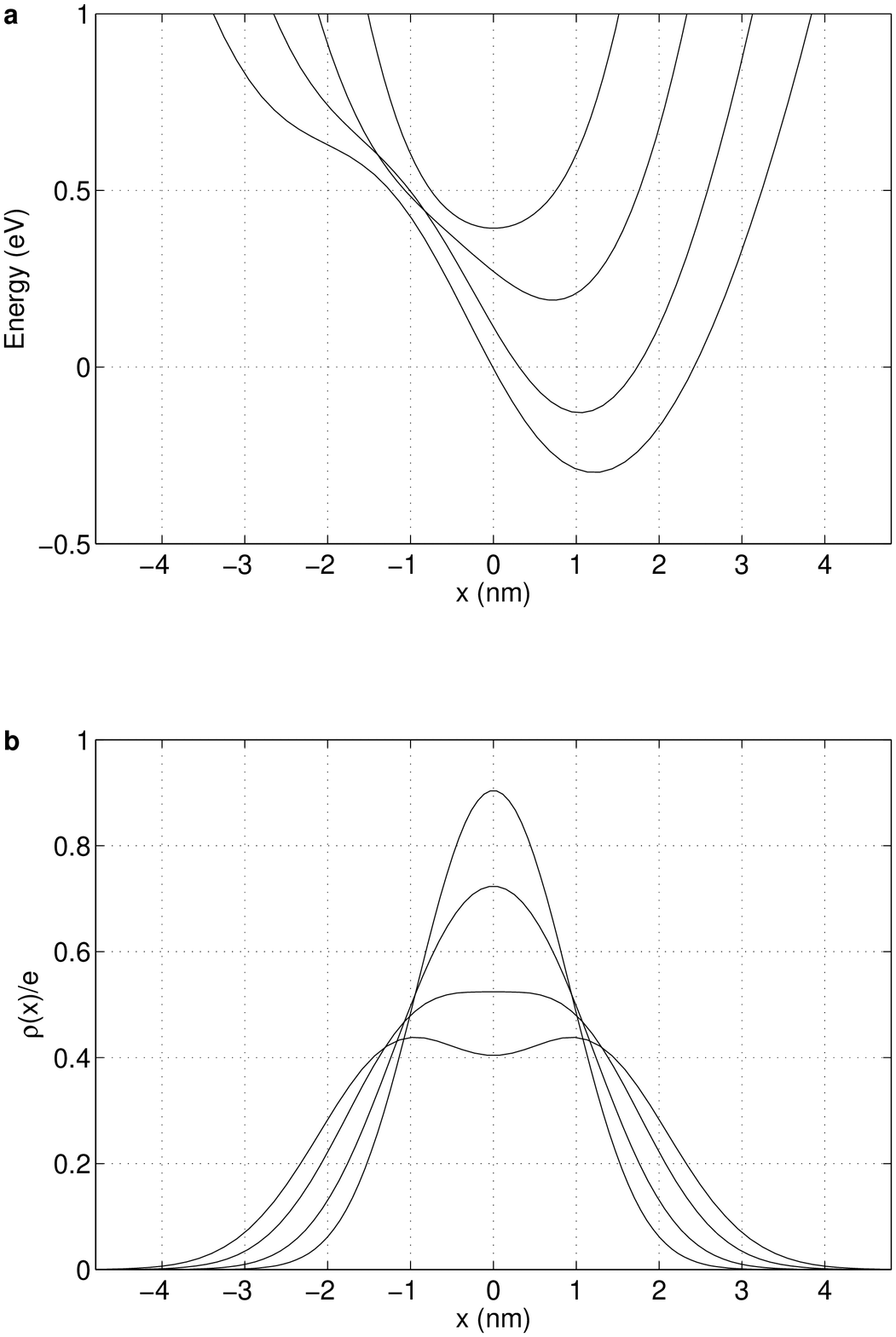}
  \end{center}
  \caption{Effective potentials and charge densities in the Hartree approximation.  The traces with center values from top and downwards are calculated using the gate widths $a=(4.8,6.4,8.0,9.6)$~nm.  The depth of the well is $V_0=1.2$~eV.  {\bf (a)} The effective potential is symmetric for the narrowest well but becomes asymmetric for wider wells.  {\bf (b)} The total electron charge density which peaks in the centre for narrow wells but shows a double peak for wider wells, reflecting charge polarisation towards away from the centre.}
  \label{fig:6}
\end{figure}

We have seen that the frozen Hartree approximation is a good approximation for weak electron correlations, as shown previously for semiconductong quantum wires \cite{Reje00,Reje03}.  In the semiconductor case the effective wire widths are at least tens of nanometers, which limits the maximum Coulomb repulsion energy when the electrons pass each other.  By contrast, the diameters of carbon nanotubes are a few nanometers at most and the Coulomb repulsion energy of the electron at the passing point is much higher, requiring a relatively deep well.  This means that the well must also be narrow in order to avoid two-electron bound states (\textit{cf} \fref{fig:2}).  Within this allowed range that gives single-electron occupation of the quantum well asymptotically, when the frozen Hartree picture breaks down, a more accurate self-consistent method, such as Hartree or Hartree-Fock, may improve the range of validity.  Although, this is strictly not valid for such an unbound system of two electrons where the probability density is dominated by one electron being far from the well, we can nevertheless gain some insight into the effects of strong Coulomb repulsion between the two electrons when they are both in the well by extending the parabolic well such that there are true bound states for both electrons and then solving Hartree equations, allowing the possibility of a different Hartree potential for each electron.  This procedure has previously been applied to two electrons in square quantum dots, giving similar results to unrestricted Hartree-Fock and showing that in the strong-correlation regime, the electrons avoid each other thereby lowering their Coulomb repulsion energy \cite{Cref00}.  These true bound-state Hartree solutions can be solved using the Hartree equations
\begin{eqnarray}
  \label{eq:16}
  \fl E\phi_n(x) = -\frac{\hbar^2}{2m^*}\frac{\rmd^2\phi_n}{\rmd x^2}
  +\left(\int_{-\infty}^{\infty}\frac{e^2}{4\pi\epsilon_0\sqrt{(x-x')^2+\lambda^2}}~|\phi_m(x')|^2~\rmd x'\right.
  \nonumber \\
  \left.+V(x)\right)\Phi_n(x),
\end{eqnarray}
where $n,m=1,2$ with $n\ne m$, and the confining potential is the potential in equation \eref{eq:7}.  The solutions are expected to give similar energies and charge densities to those of the resonant bound states when both electrons are in the well and this is indeed verified by the exact solutions.  The effective potential acting on one of the electrons in the dot is shown in \fref{fig:6}(a) with the potential of the other being the mirror image.
The parameters are the same as in \fref{fig:5}, and when $a=4.8$~nm, the two electrons, described by the two one-electron envelope functions $\phi_1(x)$ and $\phi_2(x)$, are bound tightly together in the centre by the applied potential with $\phi_1(x)=\phi_2(x)$.  In order to lower their mutual Coulomb repulsion energy, the electrons start to avoid each other as the gate width increases, which results in a breakdown of the symmetry.  The locations of the electrons can be observed in the charge density plotted in \fref{fig:6}(b).  This figure shows that the use of wider wells results in greater separation of the electrons, eventually leading to a double quantum dot with one electron occupying each well.  The energy of the bound state is positive for well parameters satisfying our requirements.  Provided that the tunnel probability through the barriers at the dot interfaces is small, the bound state energy is approximately that of the resonance state in the scattering problem.  The results give the correct picture for charge polarisation within the well but do not make any reference to the spin of the electrons.  In the strong correlation regime, the effects of spin on charge density are small but they can have a marked effect on spin-dependent resonances.  To understand this we need to solve the full two-electron problem numberically.

\section{Two-electron scattering}
\label{sec:4}

\subsection{The two-electron model}

The full two-electron Hamiltonian is, from the one-electron Hamiltonian \eref{eq:5} and the approximate mutual Coulomb interaction potential in equation \eref{eq:11},
\begin{equation}
  \label{eq:17}
  \fl H = -\frac{\hbar^2}{2m^*}\left(\frac{\partial^2}{\partial x_1^2}+\frac{\partial^2}{\partial x_2^2}\right) + V(x_1) + V(x_2)+\frac{e^2}{4\pi\epsilon_0\sqrt{(x_1-x_2)^2+\lambda^2}},
\end{equation}
where the indices label the two electrons and $V(x)$ is the effective one-electron  potential energy in equation \eref{eq:1}.  Spin-orbit coupling is negligible in carbon and has been neglected.  Hence, total spin is conserved and the eigenfunctions can, for this case of two electrons, be separated into spin and orbital factors;
\begin{equation}
  \label{eq:18}
  \Psi(x_1,\sigma_1,x_2,\sigma_2) = \Phi(x_1,x_2)\chi(\sigma_1,\sigma_2),
\end{equation}
where,
\begin{equation}
  \label{eq:19}
  \chi(\sigma_1,\sigma_2)\in \left\{
  \begin{array}{l}
    \chi_{\uparrow\uparrow}\\
    \frac{1}{\sqrt{2}}(\chi_{\uparrow\downarrow}+\chi_{\downarrow\uparrow})\\
    \chi_{\downarrow\downarrow}\\
    \frac{1}{\sqrt{2}}(\chi_{\uparrow\downarrow}-\chi_{\downarrow\uparrow})
  \end{array}
  \right.,
\end{equation}
with the first three basis states belonging to the symmetric triplet and the last state belonging to antisymmetric singlet.  To ensure overall antisymmetry with respect to interchange of spin and orbital coordinates, the orbital states must satisfy,
\begin{equation}
  \label{eq:20}
  \Phi(x_1,x_2) = \pm\Phi(x_2,x_1),
\end{equation}
with the plus sign corresponding to the singlet state and the minus sign a triplet.  The wavefunctions for these symmetric and antisymmetric orbital states are then calculated by applying the finite difference method to the envelope function equation $H\Phi = E\Phi$, after cancellation of the spin parts.  The coordinates of the two particles are mapped onto a two-dimensional square Cartesian surface which splits into two triangles about $x_1=x_2$.  We need to solve for the envelope function on only one of these triangles as the other half is always given by equation \eref{eq:20} with the sign determined by the chosen symmetry.  The triangular surface has three boundary conditions.  The hypotenuse must have either Dirichlet or Neumann boundary conditions to satisfy equation \eref{eq:20}.  The other two boundaries correspond to one electron outside the quantum dot on the source or drain side, respectively.  If the electrons are far apart, the (anti)symmetric envelope function can be replaced by factorised one-electron solutions to a good approximation and the envelope functions matched to asymptotic scattering functions,
\begin{eqnarray}
  \label{eq:21}
  \Phi(-x_0,x_2) &=& \frac{1}{\sqrt{L}}~\left(\xi_1^{\rightarrow}(-x_0)-r\xi_1^{\leftarrow}(-x_0)\right)\phi_0(x_2),
  \\
  \label{eq:22}
  \Phi(x_1,x_0) &=& \frac{t}{\sqrt{L}}~\xi_1^{\rightarrow}(x_0)\phi_0(x_1),
\end{eqnarray}
where $r$ and $t$ are the reflection and transmission amplitudes.  $\xi_1^{\leftarrow}(x)$ and $\xi_1^{\rightarrow}(x)$ are the left- and right-going asymptotic one-electron solutions, with Coulomb functions that may be approximated by plane waves.  These boundary conditions are sufficient to determine the full envelope function surface.

\subsection{Transmission and Correlation}

\begin{figure}
  \begin{center}
    \includegraphics[scale=0.5]{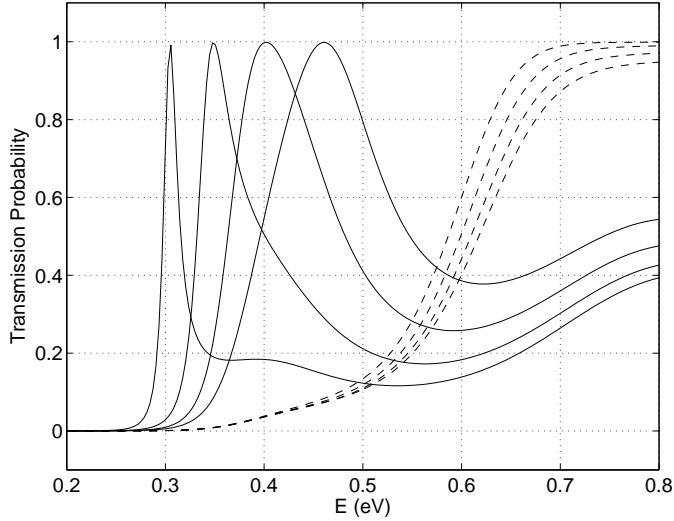}
  \end{center}
  \caption{Transmission probability as a function of the kinetic energy of the incident electron using a full two-electron calculation.  The solid and dashed curves are solutions, in which symmetric and anti-symmetric envelope functions have been enforced, respectively.  The well parameters are the same as in \fref{fig:4}.}
  \label{fig:7}
\end{figure}

\begin{figure}
  \begin{center}
    \includegraphics[scale=0.5]{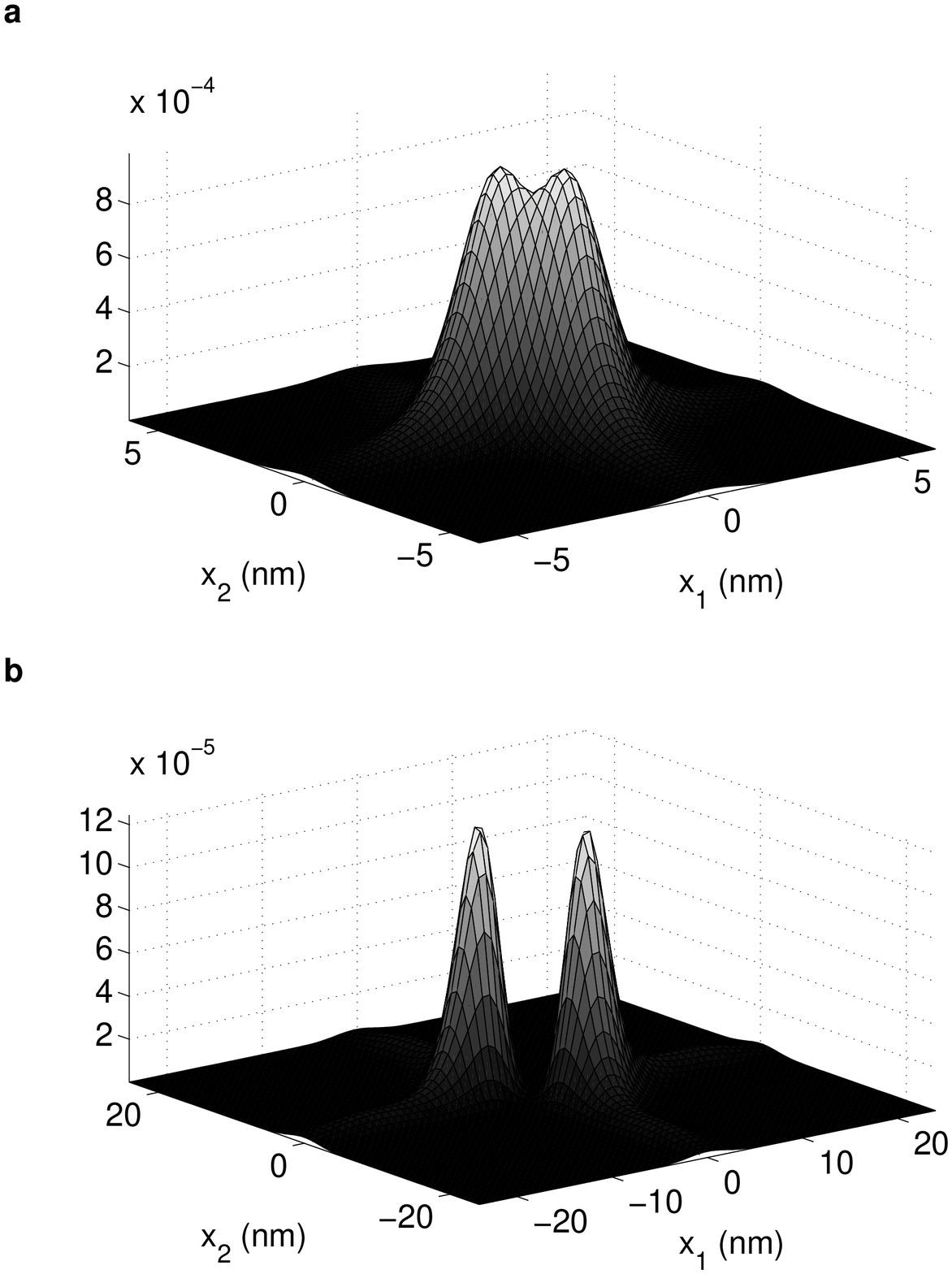}
  \end{center}
  \caption{Electron densities at the singlet-resonance.  Graph {\bf (a)} is calculated for a narrow well width of $a=4.8$~nm with $V_0=1.5$~eV and shows a typical situation where the electrons are medium correlated.  The density has started to develop a double-peak-structure, but retains a strong overlap.  The electron density in graph {\bf (b)} exhibits only a very small overlap.  This calculation has a comparably wide gate width of $a=19.2$ nm with $V_0=0.4$~eV~and illustrates a situation with strongly correlated electrons.}
  \label{fig:8}
\end{figure}

\begin{figure}
  \begin{center}
    \includegraphics[scale=0.5]{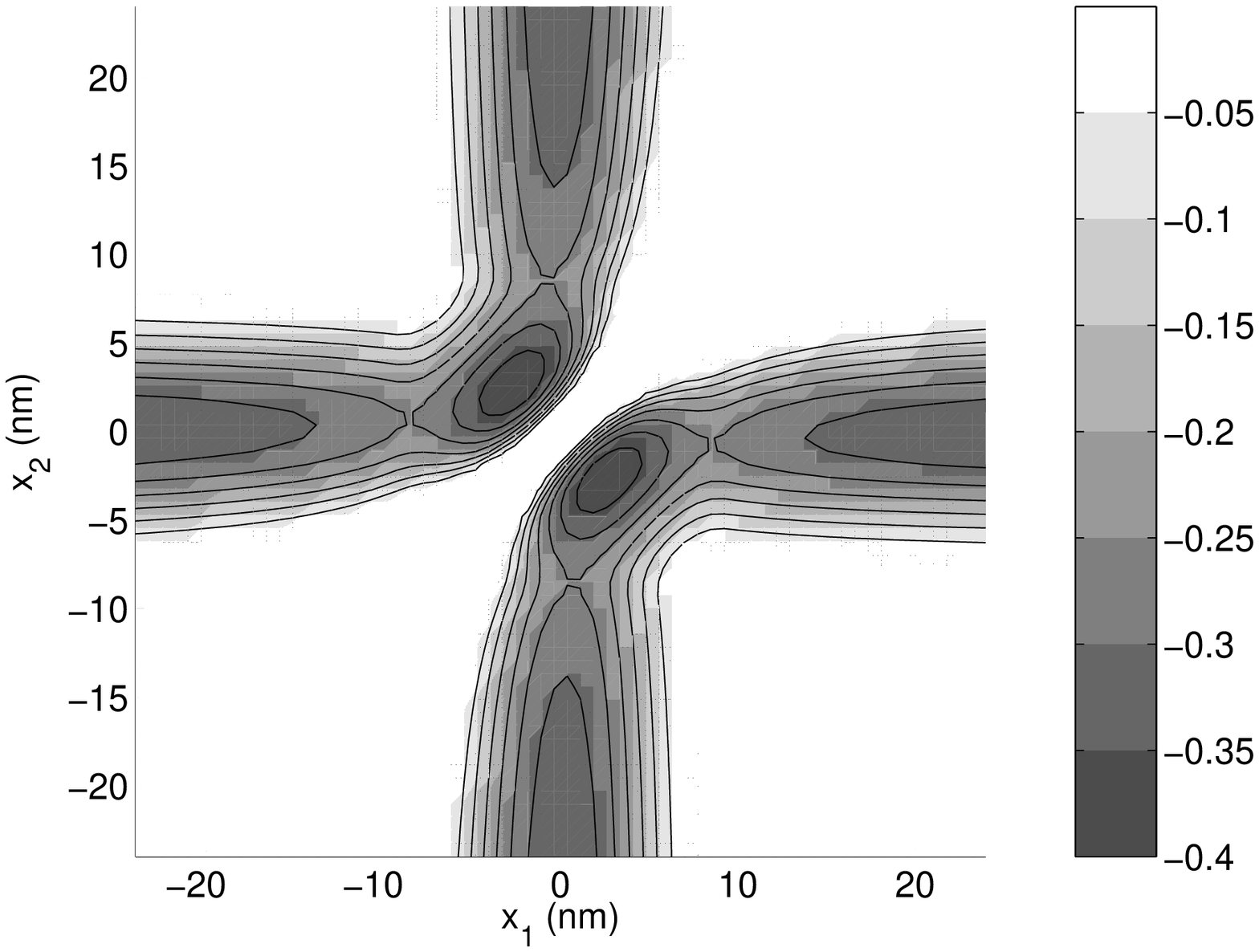}
  \end{center}
  \caption{Total two-electron potential energy with one electron at position $x_1$ and the other at $x_2$.  In the centre part of the figure, corresponding to both electrons on the quantum dot, there are two wells,implying that the electrons can avoid each other to a certain degree.  In order to better highlight the regions with low potential, all positive potential energy has been set to zero in this graph.}
  \label{fig:9}
\end{figure}

\begin{figure}
  \begin{center}
    \includegraphics[scale=0.5]{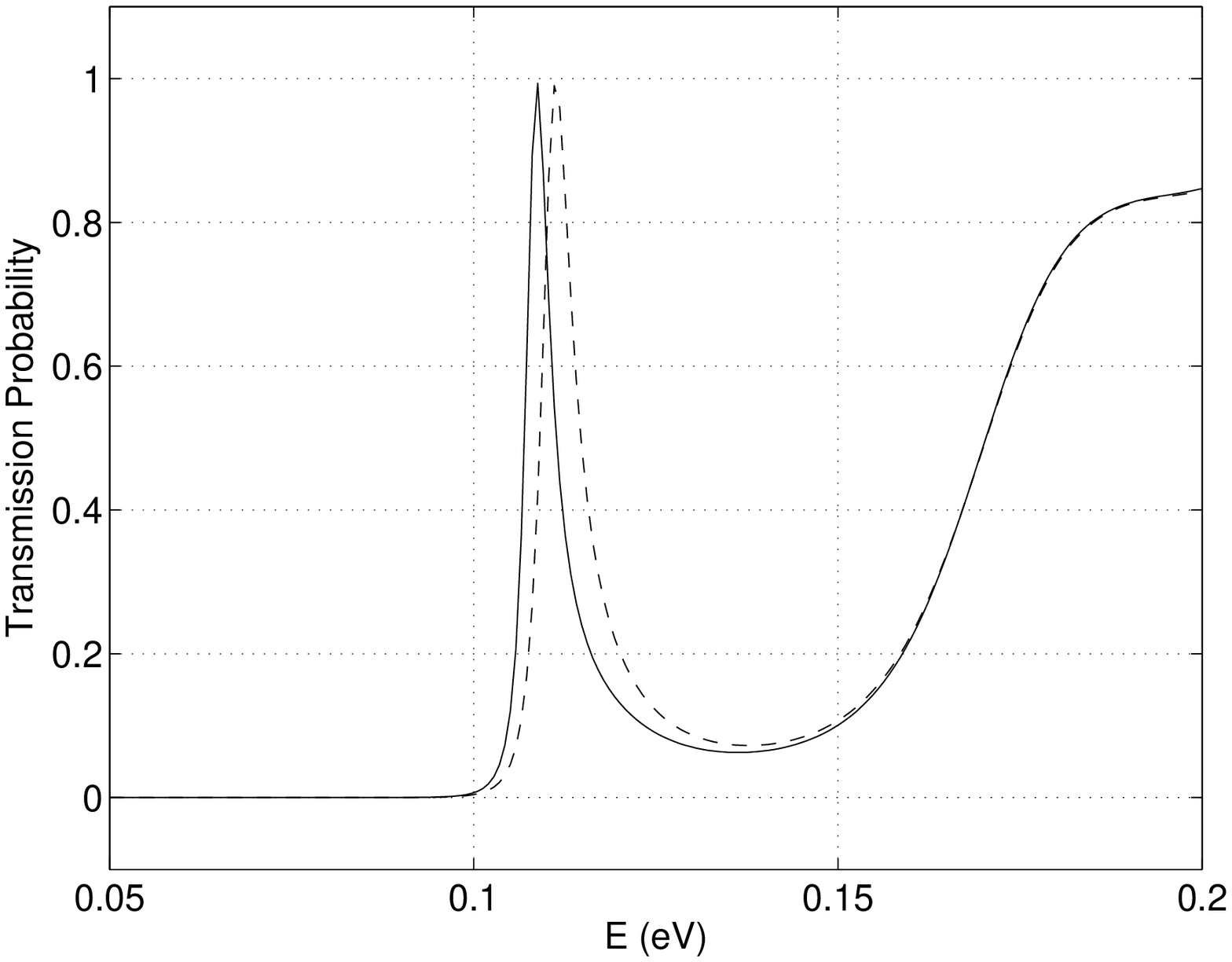}
  \end{center}
  \caption{Transmission probability for two strongly correlated electrons.  The symmetric (solid) and anti-symmetric (dashed) calculations exhibit a singlet and a triplet resonance, respectively.  The well parameters are $a=19.2$~nm and $V_0=0.4$~eV.}
  \label{fig:10}
\end{figure}

The system of linear equations is solved using the bi-conjugate gradient method.  A calculation of the transmission probability is shown in \fref{fig:7}, where we have used the same parameters as in \fref{fig:4}.
The resonances in the frozen Hartree approximation agree qualitatively with the singlet resonances (solid lines) for this very narrow well, but the resonances occur at higher energies with significantly more broadening for the highest energy resonance.  These differences are attributable to the bound electron moving away from the barriers through which the other electron is tunnelling.  As a result, the effective tunnelling barriers in the two-electron case (\textit{cf} the one-electron barriers in \fref{fig:3}) are slightly lower, and narrower at the top.  In experiments, the wide range of resonance widths may be exploited in such a way that a desired resonance width can be obtained by tuning the applied gate potential and the energy of the injected electron.  The triplet transmissions (dashed lines) in \fref{fig:7} do not exhibit resonances since the two electrons are forbidden to occupy the same quasi-bound state due to the Pauli Exclusion Principle and higher-energy single-electron states occur above the quantum dot tunnel barriers for this choice of parameters.  The electron densities of two singlet resonances are plotted in \fref{fig:8}.
The first is calculated with a well width $a=4.8$ nm.  At this width, the system is in the medium-correlated regime where the two electrons are slightly separated but maintain a sizeable overlap.  The second graph illustrates a somewhat different scenario in which the well width is $a=19.2$ nm.  The electrons are highly correlated and repel each other in the well, leading to a minimum density overlap.  In the highly-correlated electron regime, the total potential has two distinct minima as shown in \fref{fig:9}.
There are six regions with a potential energy below $-0.3$ eV.  In the outer four minima, one of the electrons is outside the dot.  In the two central minima both electrons are in the dot and are spatially separated, so that the total potential resembles a double quantum dot where each is singly occupied.  This is similar to ordinary double-dot systems (or indeed a single-dot system with two electrons in the strong correaltion regime) which have a low-lying singlet-triplet pair of energy levels.  However, in our system, these levels become resonances, or quasi-bound states as shown in \fref{fig:10}.
The separation between the resonances is small in this high-correlation example.  The overlap becomes even more pronounced for larger wells, since the resonance separation appraches zero exponentially with increasing well width.  As a consequence, antisymmetrisation of the total wavefunction becomes unnecessary.  The two high-correlation regimes with small and large resonance overlaps offer controlled entanglement.

\section{Entanglement}
\label{sec:5}

\begin{figure}
  \begin{center}
    \includegraphics[scale=0.5]{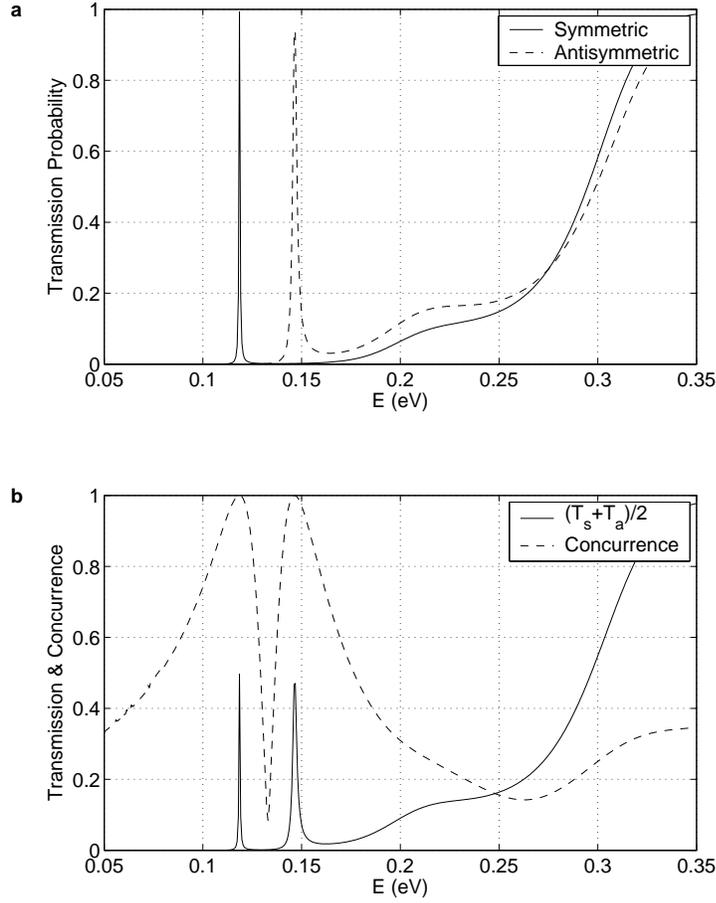}
  \end{center}
  \caption{Transmission and concurrence in the medium-correlation regime.  {\bf (a)} The symmetric and anti-symmetric calculations generate singlet and triplet resonances, which have negligible overlap.  {\bf (b)} The total transmission probability (solid) and concurrence (dashed) each have two maxima, corresponding to the singlet and triplet resonances.  The parameters are $a=12.0$~nm and $V_0=0.8$~eV.}
  \label{fig:11}
\end{figure}

\begin{figure}
  \begin{center}
    \includegraphics[scale=0.5]{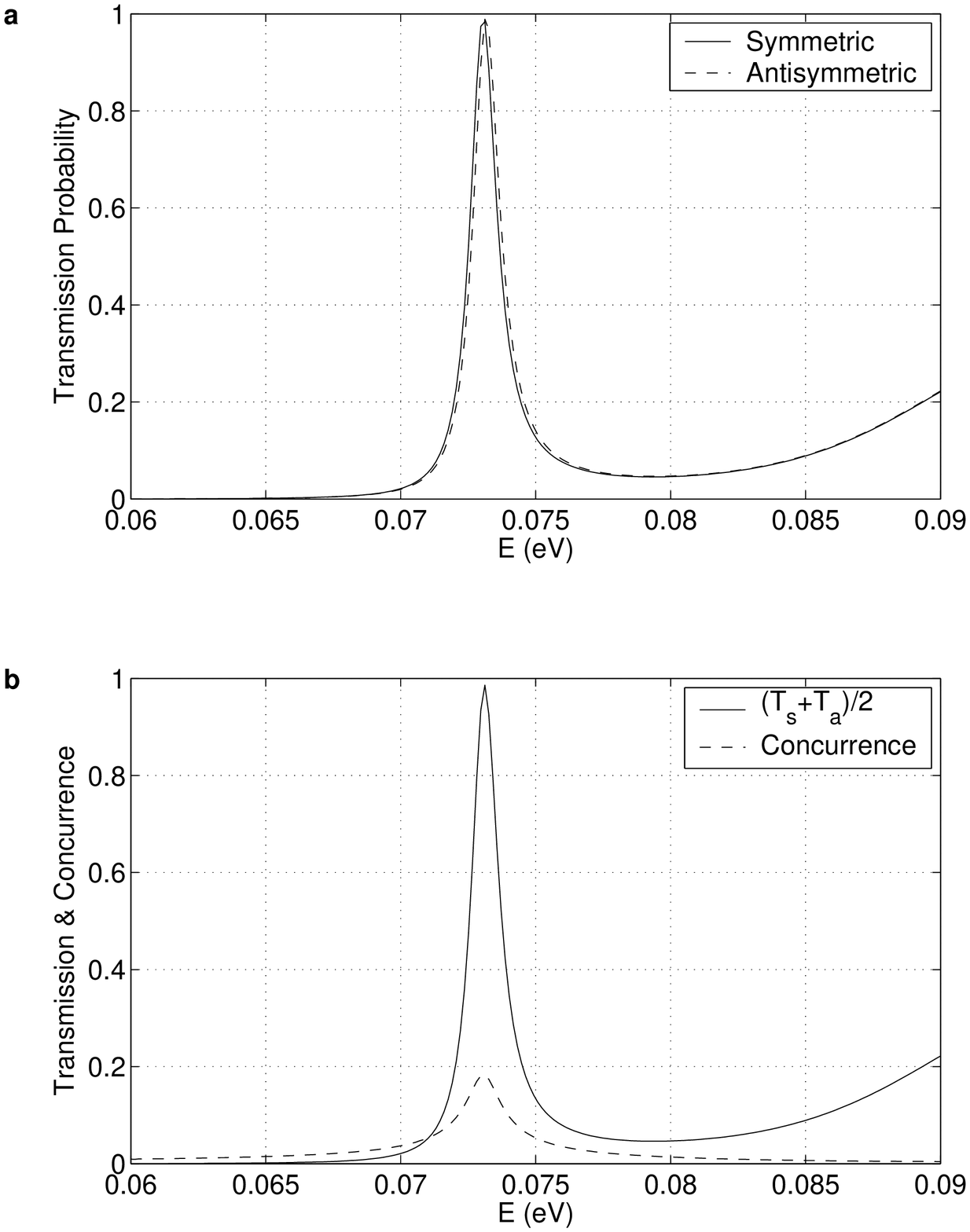}
  \end{center}
  \caption{Transmission and concurrence in the high-correlation regime.  {\bf (a)} The symmetric and antisymmetric calculations generate singlet and triplet resonances, which strongly overlap.  {\bf (b)} The total transmission probability (solid) almost reaches unity, while the concurrence (dashed) is limited.  The parameters are $a=32.0$~nm and $V_0=0.2$~eV.}
  \label{fig:12}
\end{figure}

The transmission probabilities for symmetric and antisymmetric orbital states, shown in figures \ref{fig:7} and \ref{fig:10}, correspond to singlet and triplet states respectively [\textit{cf} equations \eref{eq:18}, \eref{eq:19}, and \eref{eq:20}].  Since the Hamiltonian does not depend on spin, the total $S$ and $S_z$ must be conserved for these eigenstates.  However, if the system is prepared with the injected electron with spin-up and the bound electron with spin-down, which is not an eigenstate of the Hamiltonian, then the subsequent Coulomb interaction between the two electrons will give rise to a superposition state of the singlet and $S_z=0$ triplet.  As described in \cite{Jeff05}, the asymptotic transmitted state after scattering may be determined as follows.  The injected electron should be regarded as being in a broad wave-packet far away from the bound electron.  The initial state is a single determinant of spin-orbitals which is the sum of the singlet state state and $S_z=0$ triplet state, and cannot be factorised into a product of spin and orbital parts.  Since the scattered state is the sum of the separately evolved singlet and triplet states, then the normalised asymptotic transmitted state is
\begin{equation}
  \label{eq:23}
  \frac{1}{\sqrt{2}}~(t_s|\Psi_{0;0}\rangle+t_a|\Psi_{1;0}\rangle)=\frac{t_a-t_s}{2}|\Psi_{\downarrow\uparrow}\rangle+\frac{t_a+t_s}{2}|\Psi_{\uparrow\downarrow}\rangle,
\end{equation}
where $|\Psi_{\downarrow\uparrow}\rangle$ is a single determinant corresponding with the bound-electron again having spin-down and $|\Psi_{\uparrow\downarrow}\rangle $ the determinant in which the spins are exchanged.  Thus $(t_a-t_s)/2$ and $(t_a+t_s)/2$ are non-spin-flip and spin-flip amplitudes respectively.  The transmitted state in equation \eref{eq:23} has spin entanglement in general.  This may be quantified using concurence, which for a general pure state $|\alpha\rangle$ takes the form, \cite{Benn96,Woot98}
\begin{equation}
  \label{eq:24}
  C = \left|\frac{\langle\alpha|\sigma_y\otimes\sigma_y|\alpha^*\rangle}{\langle\alpha|\alpha\rangle}\right|,
\end{equation}
where $\sigma_y$ is a Pauli matrix and the operator inside the bra-ket is a total spin-flip operator.  The initial state $|\Psi_{\downarrow\uparrow}\rangle$ has no concurrence.  The concurrence of the final state, however, is
\begin{equation}
  \label{eq:25}
  C = \frac{|t_s^2-t_a^2|}{|t_s|^2+|t_a|^2}.
\end{equation}
This gives maximum entanglement when $T_s\equiv |t_s|^2=1$ and $T_a\equiv |t_a|^2=0$ or vice versa.  The concurrence is plotted together with the transmission spectra in \fref{fig:11}.
The parameters of the applied potential in the well, $a=12.0$~nm and $V_0=0.8$~eV, were chosen so that the singlet and triplet resonances have a negligible overlap, and therefore the concurrence has maxima at the resonances.  Between the resonances, there is a minimum in the concurrence when $t_s^2\approx t_a^2$.  This condition is less likely to become satisfied outside the resonances due to the unbalance between the transmission probabilities, and therefore the entanglement decays slower than between the resonances.  The well isolated pair of singlet-triplet resonances suggests that the Hamiltonian can be approximated by an effective Heisenberg spin Hamitonian
\begin{equation}
  \label{eq:26}
  H = J~S_1\cdot S_2,
\end{equation}
where $J$ is the exchange constant.  To reproduce the resonance separation,  $J = E_T-E_S$, which is approximately $J\approx 30$ meV in \fref{fig:11}.  This may be regarded as a generalisation of the Heisenberg model for the low-energy spin-triplet of two electrons fully confined in a quantum dot \cite{Jeff96} to the present case of resonant bound states.  This generalisation is valid only when resonance widths are much less than the energy separation, which is the case in \fref{fig:11}.
A different concurrence picture arises when the singlet and triplet have a large overlap, which is seen by the results in \fref{fig:12}.
The total well width here is $a=32$~nm.  The two electrons are now so far apart that the symmetric and antisymmetric wavefunctions can scarcely be distinguished, whereas the resonance widths are still relatively large, leading to a large overlap between the resonances.  This has two main consequences.  First, the transmission probability of the initial state can reach almost unity instead of $50\%$ as in the previous example.  Second, the system can only acquire partial entanglement.  Both consequences are confirmed by \fref{fig:12}.  The maximum concurrence in this regime can be estimated by assuming that the transmission coefficients for the symmetric and antisymmetric solutions are the same.  The amplitudes can then be expressed as $t_s\approx |t|e^{i(\phi-\theta)}$ and $t_a\approx |t|e^{i(\phi+\theta)}$.  The phase difference is estimated to be $2\theta\approx (E_T-E_S)\tau/\hbar$ where $\tau=\hbar/\Gamma$ is the lifetime of the quasi-bound state with resonance width $\Gamma$.  Thus, the maximum concurrence in this highly correlated regime is approximately
\begin{equation}
  \label{eq:27}
  C_{max} \approx \left|\sin \frac{E_T-E_S}{\Gamma}\right|\approx \left|\frac{E_T-E_S}{\Gamma}\right|,
\end{equation}
which is an excellent approximation to the concurrence in \fref{fig:12}, which was calculated using the exact expression of equation \eref{eq:25}.

After the entanglement process the spin of the propagating electron could in principle be measured using a spin filter, such as a Zeeman-split quantum dot produced by gates along the nanotube, with subsequent charge detection using a single-electron transistor.  Measuring the static spin directly would be more difficult. However, this may be verified indirectly as follows \cite{Pope}.  After detection of the spin of the flying qubit the static qubit should have opposite spin.  This may then be verified by injecting a further spin of the same polarity as the (now known) static spin and subsequently measuring it using a spin filter.

\section{Conclusion}
\label{sec:6}

Spin-entanglement can be created between a propagating electron and a bound electron in a single-walled nanotube, provided that neither the bound electron is ionised nor the propagating electron is captured by the well.  These requirements lead to conditions on the well.  We have shown that when these conditions are satisfied, the model system may be classified into three  regimes.  For very narrow wells where $a \le 6$~nm, the effective potential exhibits a double barrier structure with a single well.  In this regime, there is a singlet resonance but not a triplet resonance.  For intermediate, $6$~nm $< a \le 20$~nm, and large, $a> 20$~nm, well widths, the potential exhibits a double-well structure, with both singlet and triplet resonances.  In the intermediate regime the energy separation is of order meV and larger than the width of the resonances.  In this regime the singlet-triplet resonances may be mapped onto a Heisenberg model with an antiferromagnetic exchange constant $J$ which may be tuned by the well parameters.  When the resonances are resolvable, maximum entanglement can be generated with a transmission probability of approximately $50\%$ for separable initial states.  This is because the initial separable state may be written as an equal superposition of fully-entangled singlet and triplet states and on resonance only one component is transmitted, to a good approximation.  For wide wells, on the other hand, full transmission can be achieved but the overlapping resonances give only small entanglement, approximately equal to the singlet-triplet energy separation in units of linewidth.

\ack
This research is part of the QIP IRC (GR/S82176/01, is supported through the Foresight LINK Award Nanoelectronics at the Quantum Edge by EPSRC (GR/R660029/01, Hitachi Europe Ltd.  GADB thanks EPSRC for a Professorial Research Fellowship (GR/S15808/01).  JHJ acknowledges support from the UK MOD.

\end{document}